\RequirePackage[]{graphicx} 
\graphicspath{{./Figures/}{./Graphs/}{./Drawings/}}

\documentclass{mrmsub}

\usepackage[utf8]{inputenc}
\usepackage{graphicx}
\usepackage{environ}
\usepackage{listings}
\usepackage{booktabs}
\usepackage{blindtext}
\usepackage{url}
\usepackage{siunitx}
\usepackage{caption}
\usepackage{pdflscape}

\usepackage[english]{babel}
\usepackage{csquotes}
\usepackage[
  backend=biber,
  natbib = true,
  style=ext-numeric-comp,
  sorting=none,
  minbibnames=3, maxbibnames=6,
  uniquename=false, uniquelist=false,
  giveninits=true, terseinits,
  articlein=false, innamebeforetitle=true,
  isbn=false,
  urldate=long, dateabbrev=false,
  autocite=superscript,
]{biblatex}

\DeclareNameAlias{sortname}{family-given}

\DeclareNameAlias{author}{sortname}
\DeclareNameAlias{editor}{sortname}
\DeclareNameAlias{translator}{sortname}

\DeclareDelimAlias{finalnamedelim}{multinamedelim}

\DeclareFieldFormat
  [article,inbook,incollection,inproceedings,patent,thesis,unpublished]
  {title}{#1}

\renewbibmacro*{journal+issuetitle}{%
  \usebibmacro{journal}%
  \setunit*{\addspace}%
  \iffieldundef{series}
    {}
    {\newunit
     \printfield{series}%
     \setunit{\addspace}}%
  \usebibmacro{issue+date}%
  \setunit{\addsemicolon\addspace}%
  \usebibmacro{volume+number+eid}%
  \setunit{\addcolon\space}%
  \usebibmacro{issue}}

\DeclareFieldFormat[article,periodical]{number}{\mkbibparens{#1}}

\renewbibmacro*{issue+date}{%
  \usebibmacro{date}}

\renewbibmacro*{pubinstorg+location+date}[1]{%
  \printlist{location}%
  \iflistundef{#1}
    {\setunit*{\addsemicolon\space}}
    {\setunit*{\addcolon\space}}%
  \printlist{#1}%
  \setunit*{\addsemicolon\space}%
  \usebibmacro{date}%
  \newunit}

\DeclareFieldFormat{pages}{#1}

\DeclareFieldFormat{doi}{%
  doi\addcolon\space
  \ifhyperref
    {\href{https://doi.org/#1}{\nolinkurl{#1}}}
    {\nolinkurl{#1}}}

\DeclareFieldFormat{language}{}
\DeclareFieldFormat{origlanguage}{}%

\AtEveryBibitem{%
\clearlist{language}%
\clearfield{number}%
\clearfield{month}%
\clearfield{day}%
\clearfield{urlyear}%
\clearfield{urlmonth}%
\clearfield{issn}%
\clearfield{series}%
\clearfield{pagetotal}%
\clearfield{note}%
\clearfield{language}%
\clearfield{origlanguage}%
\clearfield{langid}%
\clearfield{eprintclass}%
\ifentrytype{article}{
	\iffieldundef{volume}{}{\clearfield{doi}}%
	\clearfield{eprint}%
	\clearfield{url}%
    \clearfield{publisher}%
}{}%
\ifentrytype{book}{%
	\clearfield{eprint}%
	\clearfield{doi}%
	\clearfield{url}%
}{
	\clearname{editor}%
}
\ifentrytype{inproceedings}{%
	\clearfield{eprint}%
	\clearfield{url}%
	\clearfield{doi}%
	\clearfield{publisher}%
	\clearlist{publisher}%
	\iffieldundef{booktitle}{}{\clearfield{eventtitle}}%
}{}%
\iffieldundef{eprint}{}{\clearfield{url}}
\iffieldundef{doi}{}{\clearfield{url}}
}

\DeclareSIUnit{\pixel}{px}

\newcommand{\figref}[2][]{\hyperref[#2]{Figure~\ref*{#2}#1}}
\newcommand{\tblref}[2][]{\hyperref[#2]{Table~\ref*{#2}#1}}
\renewcommand{\eqref}[1]{\hyperref[#1]{Eq.~\ref*{#1}}}

\providecommand{\doi}{}
\renewcommand{\doi}[1]{\href{https://doi.org/#1}{#1}}

\newcommand{\preSubmission}{}
\newcommand{\wordCountAbstract}{230}
\newcommand{\wordCountBody}{3800}

\providecommand{\del}[2][]{}
\providecommand{\dels}[2][]{}
\providecommand{\printfunding}{}

\newcommand{\absdiv}[1]{%
	\par\addvspace{.5\baselineskip}
	\noindent\textbf{#1}\quad\ignorespaces
}

\title{Improving Multi-Delay-ASL through specialized reconstruction}

\corres{Martin Uecker, Graz University of Technology,
Institute of Biomedical Imaging, Stremayrgasse~16/3, 8010~Graz, AUSTRIA, \email{uecker@tugraz.at}}

\author[1]{Ingmar Sorgenfrei}
\author[2]{Qinyang Shou}
\author[1]{Ingrid Barth}
\author[1]{Martin Uecker}
\author[2]{Danny JJ Wang}
\author[1]{Rudolf Stollberger}
\authormark{I. Sorgenfrei \textit{et al.}}

\address[1]{Institute of Biomedical Imaging, TU Graz, Austria}
\address[2]{Laboratory of Functional MRI Technology, Mark \& Mary Stevens Neuroimaging \& Informatics Institute, University of Southern California, USA}

\keywords{diffusion posterior sampling, MRI, image reconstruction, parallel imaging, Bayesian reconstruction}

\articletype{Journal Article}

\finfo{
The data collection of this research was funded by US National Institutes of Health grant R01EB028297.
}

\abstract{
\absdiv{Purpose} Although image reconstruction has received relatively little attention in ASL research to date, it has the potential to address several challenges in ASL. As well as speeding up measurements by increasing undersampling and reducing measurement artifacts, it can improve the signal-to-noise ratio (SNR) of a given data set and the reproducibility of examinations. 
\absdiv{Methods} This work focuses on extending a dedicated ASL reconstruction approach (ASL-TGV) \cite{Spann2020} to multi-delay data and applying it to a high-resolution pCASL test-retest dataset. To show not only improvement in the Perfusion Weighted Images (PWIs), Cerebral Blood Flow and Arterial Transit Time was estimated and the test-retest reliability was estimated using the within subject Coefficient of Variance (wsCV), the Intraclass Correlation Coefficient (ICC) and Root-Mean-Squared-Error (RMSE). 
\absdiv{Results} The Perfusion-Weighted-Images reconstructed from highly undersampled single-shot data using ASL-TGV are clearly improved compared to a fully-sampled reference from the same data even after denoising, especially for long PLDs and the outermost slices. 
SNR calculated in grey and white matter ROIs shows an improvement of 62\% and 35\% respectively. 
The CBF maps produced from the ASL-TGV images have an improved test-retest reliability. 
\absdiv{Conclusion} ASL-TGV, which is now implemented in the Berkeley Advanced Reconstruction Toolbox (BART), can be used with any type of ASL labeling or data acquisition and any existing image post-processing pipeline and can improve SNR for the PWIs and reproducibility of the CBF maps. 
\absdiv{Keywords} Arterial Spin Labeling, Image Reconstruction, Reproducibility
}

\begin{document}

\maketitle

\section{Introduction}\label{sec_intro}
Arterial Spin Labeling (ASL) \cite{Detre1992} is a non-invasive magnetic resonance imaging (MRI) technique that delivers perfusion information without exogenous contrast agents. 
It has been shown to be a valuable tool in diagnosing multiple perfusion-related disorders from stroke to neurodegenerative diseases \cite{Telischak2015,Haller2016,Lindner2023}. 
The use of blood as an endogenous tracer enables longitudinal studies and avoids the contraindications due to contrast agents in children or people suffering from kidney disease. 
Despite the possibilities and advantages of ASL, it remains a technically challenging modality. 
The signal-to-noise ratio (SNR) is inherently low and the labeling and control conditions before each acquisition lead to long scan times\cite{Alsop2015}. 
Additionally, while it has been shown that the perfusion values derived from Perfusion-Weighted Images (PWI) are reasonably reproducible \cite{Parkes2004,Petersen2010,Gevers2011,Chen2011_testretest,Jain2012,Mutsaerts2014,Mutsaerts2015,Cohen2020}, even when doing voxel-wise comparisons \cite{Ssali2016}, perfusion proves to be one of the least reproducible quantitative parameters in MRI \cite{McGuire2017,Melzer2020}

The field of ASL is shaped by the consensus recommendations provided by Alsop et al \cite{Alsop2015} for single-delay ASL (SD-ASL).
More recently, a series of reviews, which cover clinical applications of ASL \cite{Lindner2023}, a presentation of community-developed ASL pipelines \cite{Fan2023}, updated guidelines for acquisition of multi-delay ASL (MD-ASL)\cite{Woods2024} and a review of recent technical developments \cite{Hernandez-Garcia2022} have reflected the substantial progress in development of data acquisition and image processing. 
However, the intermediate step between the two, image reconstruction, has remained comparatively under-explored in the ASL community. 
This is notable, as advanced image reconstruction has been shown to address some inherent problems in ASL and is particularly suited to ASL data. 

ASL data acquisition, where averages of the same image or similar images of different Post-Labeling Delays (PLDs) are acquired multiple times, inherently leads to large amounts of redundant information in the data. 
This can be exploited by specialized image reconstructions, which can improve image quality, reproducibility and allow higher undersampling factors, which can shorten scan times. 
Current recommendations, which do not take advanced reconstruction into account, limit undersampling in ASL to 2-3 \cite{Alsop2015,Hernandez-Garcia2022} because of the associated SNR penalty for the already low-SNR method. 
Specialized reconstructions can overcome that without interfering with the established acquisition or image post-processing pipelines. 

Several approaches for advanced image reconstruction in ASL have been introduced. Looking at examples of reconstructing brain PWIs, there are reconstructions based on modeling a dictionary representation of the signal \cite{Zhao2015}, fingerprinting \cite{Wright2018}, regularization on the PWI \cite{Mehranian2020} or deep learning \cite{Gong2022}. 
Interestingly, these results have featured little if at all in the updated recommendations for ASL \cite{Woods2024} or the review of the state of the art \cite{Hernandez-Garcia2022}. 
Additionally, all major ASL processing pipelines presented in Fan et al\cite{Fan2023} expect already reconstructed images and there is no equivalent collective effort to collect specialized ASL image reconstruction software. 

A powerful, specialized SD-ASL reconstruction was introduced by Spann et al \cite{Spann2020}. 
Contrary to some of the approaches above, it does not require any assumption on the perfusion model and can - without alteration - be applied to any type of labeling and sampling. 
It applies spatio-temporal regularization on the control-label pairs and the PWIs concurrently and combines that with a time-dependent CAIPIRINHA undersampling pattern. This enables 3D single-shot acquisitions with reasonable echo train lengths using acceleration factors up to six. 
These accelerated, single-shot acquisitions also inherently enhance motion robustness, proving the potential of advanced image reconstruction to be a valuable tool in the ASL processing chain. 

In this work, we extend this framework to MD-ASL, which enables estimation of both CBF and ATT from highly undersampled, single-shot data. 
It will be shown that reconstruction of this undersampled data is impossible without advanced reconstruction and how both SNR of the images and reproducibility of the perfusion values improve. 
To make adaptation in the ASL community as frictionless as possible, all methods are implemented in the open-source Berkeley Advanced Reconstruction Toolbox (BART) \cite{BART}, which makes the presented work available to be reproduced, used and expanded upon.   

\section{Theory \& Methods}\label{sec_methods}

The ASL reconstruction framework presented in this work builds on Spann et al \cite{Spann2020}.
Their specialized ASL reconstruction together with a few relevant concepts will be briefly re-introduced before the extension to MD-ASL is discussed. For a full introduction to the ASL reconstruction, refer to \cite{Spann2020}. 

\subsection{Single Delay ASL Reconstruction}

A generalized expression for a regularized reconstruction problem in MRI can be given as 
\begin{align}
	u^* \in \text{argmin}_u \,\,\frac{1}{2}||K(u) - d||_2^2+R(u)\label{eq:gen_recon}
\end{align}
where the desired image $u^* \in \mathbb{C}^{N_x\times N_y \times N_z \times N_t}$ is estimated from noisy k-space data $d \in \mathbb{C}^{N_x\times N_y \times N_z \times N_c \times N_t}$ using the forward operator $K$ and a regularization term $R$. $N_x\times N_y \times N_z$ define the size of the spatial dimensions, $N_t$ the number of frames acquired over time and $N_c$ the number of acquisition coils. In MRI reconstruction, $K$ typically includes application of the coil sensitivities, the Fourier transform and the sampling pattern. It is included in the data fidelity term, which enforces consistency between the acquired data $d$ and the reconstructed image $u$. $R$ can be one or more regularization terms. 

The regularization used in this work is Total Generalized Variation of second order (TGV)\cite{Knoll2011} that enforces a piece-wise smooth image and has the same edge-preserving capabilities of Total Variation (TV)\cite{Rudin1992,Block2007}. TGV is defined as follows:
\begin{equation}
	\text{TGV}(u) = \min_v \alpha_1 |\nabla_\beta (u) - v|_1 + \alpha_0 |\epsilon_\beta v|_1
\end{equation}
$\nabla_\beta=(\beta_x \partial_x, \beta_y \partial_y, \beta_z\partial_z, \beta_t\partial_t)^T$ is a weighted derivative in all spatial and the temporal dimensions. $\epsilon_\beta=(\nabla_\beta + \nabla_\beta^T)$ is the symmetrized derivative. The weights are summarized in the variable $\beta$. This allows including prior knowledge about the sparsity in the data along different dimensions. 

The generic reconstruction equation \ref{eq:gen_recon} was adapted for ASL by Spann et al \cite{Spann2020}. 
\begin{align}
	(c^*,l^*) \in\, &\text{argmin}_{(c,l)}\,\,\frac{1}{2}||K(c,l) - d_{c,l}||_2^2 + \notag\\ 
	&\lambda (\theta_1 \text{TGV}(c,l) + \theta_2 \text{TGV}(c-l))\label{eq:asl_recon}
\end{align}
$(c,l)$ refers to the control and label images and $d_{c,l}$ is the full data set of both the control and label data. The data fidelity term and one TGV term is applied to the control and label images concurrently. The addition of another TGV term applied to the difference image, the PWI, ensures that the regularization does not suppress or distort the perfusion information \cite{Spann2020}. $\lambda$ governs the strength of the overall regularization, while $\theta_1$ and $\theta_2$ balance the strength of the TGV terms on the control and label images versus the PWIs. This reconstruction method will from now on be referred to as ASL-TGV. 

It must be noted that the notation of ASL-TGV differs from the one used by Spann et al\cite{Spann2020}. While the formulation is functionally identical, the notation here reflects the way the reconstruction was newly implemented into BART. 
The previous implementation is in the AVIONIC toolbox \cite{Schloegl2017}.

\subsection{Extension to Multi-Delay Data}
\begin{center}
	\begin{figure}
		\includegraphics[width=400pt]{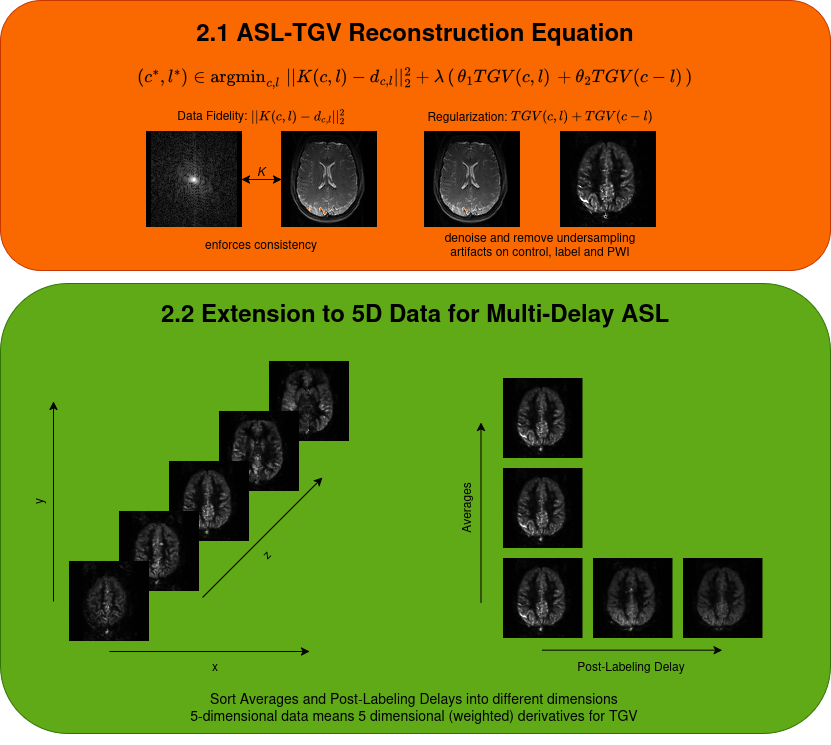}
		\caption{This figure is a visual representation of the two sections in the theory chapter. The top part of the graph reintroduces the ASL-TGV reconstruction from Spann et al \cite{Spann2020}. The meaning of the two parts of the reconstruction problem, data fidelity and regularization, are exemplified. The data fidelity term enforces that the final image represents the original k-space data. They are connected via the forward operator $K$. The regularization is applied to both the control-label images and the PWI, which is the crucial feature to ensure that the perfusion information is not disturbed during regularization. The lower section is an overview of the extension to 5-dimensional data. There are three spatial dimension and instead of a single time dimension, there are two. They separate the multiple acquisitions at a specific PLD, which are called averages, and the acquisitions at different PLDs.  \label{fig_Theory}}
	\end{figure}
\end{center}

The above reconstruction equation was originally formulated for SD-ASL. 
It was designed to  reconstruct a single 2D or 3D image or multiple averages of 2D or 3D ASL images. 
To account for signal changes across different PLDs in MD-ASL, an additional data dimension is added and will be used for regularization. 
Thus, there is one dimension where the multiple acquisitions at each PLD are sorted into, which will be called averages, and another one for the multiple PLDs. 
This changes the above definitions to $c,l \in \mathbb{C}^{N_x\times N_y \times N_z \times N_d \times N_a}$ and $d_{c,l} \in \mathbb{C}^{N_x\times N_y \times N_z \times N_c\times N_d \times N_a}$, where $N_t$ has been replaced with $N_d\times N_a$, which correspond to the size of the delay-dimension, the different PLDs, and the average-dimension, the multiple acquisitions at each specific PLD. 

Consequently, the gradients for TGV must be expanded to a fifth dimension together with the scaling factor $\beta$. This enables different regularization strength along the averages versus the delays. 

The different parts of the reconstruction equation and the 5-dimensional image are visualized in figure \ref{fig_Theory}.

A similar strategy of separating consecutive acquisitions into different data dimensions according to their meaning has been successfully applied previously by, e.g., Feng et al\cite{Feng2016}.  
\section{Data Acquisition \& Processing}
\subsection{Acquisition}
\begin{center}
	\begin{figure}
		\includegraphics[width=400pt]{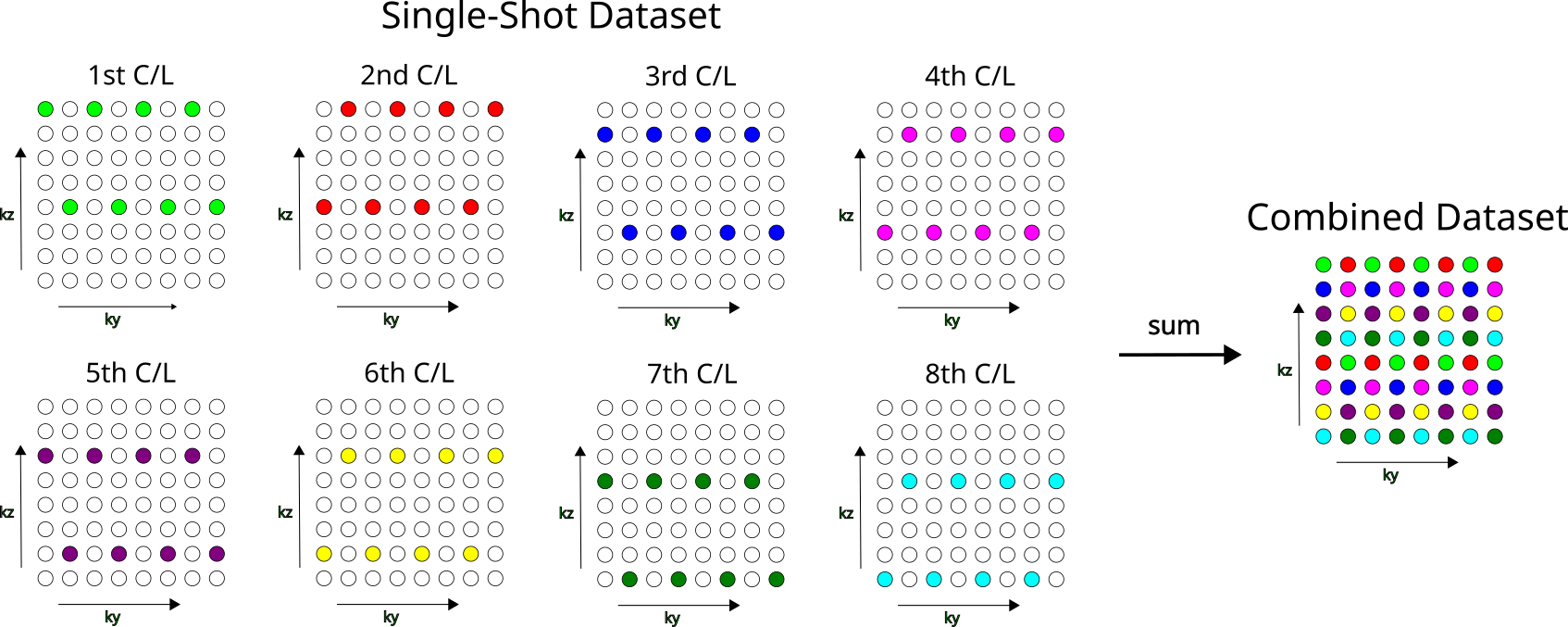}
		\caption{The CAIPI undersampling pattern, which is used to acquire one full k-space in 8 segments for each Control-Label pair. When each segment is considered as an undersampled k-space from which an image is reconstructed, the dataset is referred to as the single-shot dataset (SSD in following figures). When the 8 acquisitions are summed up into one fully sampled k-space, it is referred to as the combined dataset (CD in following figures) \label{fig_caipi}}
	\end{figure}
\end{center}

To evaluate ASL-TGV, a dataset previously used by Shou et al \cite{Shou2025} was employed. The dataset contains high resolution MD-pCASL data with an additional T1w image acquired on a 3T MR scanner (Prisma, Siemens Healthcare, Germany) using a 32-channel head coil. Each of the 21 healthy, pediatric subjects (age $=13\pm2.5$ years, 13 males, 8 females) was scanned twice with two weeks between measurements allowing evaluation of the test-retest reliability of the reconstruction. The time-dependent CAIPI undersampling pattern of the ASL sequence depicted in figure \ref{fig_caipi} covers the entire k-space in 8 segments, but each segment can also be used as an undersampled, single-shot acquisition of the k-space. 

The full imaging protocol is as follows: TR = 6180 ms, TE = 52.5 ms, FOV = $192\times 192\times 96$ mm$^3$, resolution of $2\times2\times2$ mm$^3$, matrix size of $96\times96\times48$, Labeling Duration (LD) = 1500 ms, Post-Labeling Delays (PLDs) = (600,1000,1400,1800,2200) ms. One fully sampled control and label image per PLD was acquired, each of them in 8 segments, which makes 80 acquisitions in total. This translates to 8 averages per PLD, if each segment is considered as a single-shot acquisition. This dataset, when all acquired segments are combined for fully sampled k-spaces, will be named the combined dataset (CD in figures) and has size $96\times96\times48\times32\times5\in \mathbb{C}^{N_x\times N_y \times N_z \times N_c \times N_d}$. This dataset, when each segment is used individually, will be named the single-shot dataset (SSD in figures) with size $96\times96\times48\times32\times5\times8\in \mathbb{C}^{N_x\times N_y \times N_z \times N_c\times N_d \times N_a}$.

\subsection{Reconstruction}
For the reconstruction, the ASL k-space data is converted into the BART format using Matlab (R2023b, The MathWorks, Natick, MA). 
The following steps are all completed within BART. 
First, the same coil compression is applied to both datasets, reducing the data size to 6 virtual coils.
The combined dataset is reconstructed using a Fast Fourier Transform (FFT) combined with a root-sum-of-squares (RSS) over the coil dimension. 
This reconstruction will be referred to as the FFT reconstruction. 
For the single-shot dataset, first, the coil sensitivities are estimated using nonlinear inversion\cite{Uecker2008} from the data averaged over all control-label images. 
Then, the ASL-TGV reconstruction is made using $\lambda=0.4$, $\theta=(1,5)$, $\beta=(1,1,1,1,10)$, $\alpha_1=1$, $\alpha_0=\sqrt{3}$, which is solved using Alternating Direction Method of Multipliers (ADMM) in 500 iterations. 
The data is inversely scaled with $w=1.7\times10^{-7}$ before the reconstruction and scaled back afterwards. 
$\theta$ and $\beta$ were optimized using simulated data. 
The value estimated for $\lambda$ from simulations was not optimal due to the simplistic nature of the simulations. 
Thus, the optimal value of $\lambda$ was estimated using the L-curve heuristic \cite{Hansen1999} on the in-vivo data. 
This reconstruction will be referred to as ASL-TGV. For a detailed explanation how to do this reconstruction with BART, please refer to the appendix \ref{appBARTrecon}. 

The M$_0$ image is reconstructed with a SENSE reconstruction in BART using the identical coil compression and sensitivities as the ASL data. 
For the T1w image, the images provided by the scanner were used. All BART reconstructions and the sensitivity calculation were run on an NVIDIA H100 graphics card. 

As a comparison, the single-shot dataset is also reconstructed once using a Conjugate-Gradient SENSE reconstruction (CG-SENSE) \cite{Pruessmann2001} using the same coil sensitivities as the ASL-TGV reconstruction.

\subsection{Processing}
The images are denoised using an adaptive Wiener Filter \cite{Wells2010} with the global noise estimate calculated according to Donoho and Johnstone \cite{Donoho1994}. CBF and ATT are estimated from the denoised images using the general kinetic model \cite{Buxton1998} with custom Matlab code. Brain, grey matter and white matter masks are created with FSL \cite{FSL_gen_Jenkinson2012} to use during fitting and for further tissue-specific analysis. The grey matter mask are the voxels with a minimum estimate of 80\% grey matter content and for the white matter mask the cutoff is a minimum of 90\%.  

\subsection{Analysis}
As a first step, visual comparisons of all the reconstructed images were made and the effect of denoising on the images is discussed. 
Then, excluding the images from the CG-SENSE reconstruction, SNR of the PWIs was estimated in the grey and white matter using two cuboid regions of interest (ROIs) multiplied by the grey or white matter mask created by FSL. 
The size of the cuboid was $15\times10\times5$ voxels for white matter and $20\times10\times5$ voxels for grey matter and was placed around the center slice. 
The ROIs are depicted in figure \ref{fig_SNR}. 
The mean intensity of grey or white matter pixels within the ROIs was taken as signal strength and the standard deviation within those same pixels was taken as a measure of noise. 
The SNR was then linearly defined as signal divided by noise. 
The CBF maps from the different reconstructions were also visually compared and the average grey and white matter perfusion were compared to literature. 

Before assessing the test-retest reliability, the dataset was checked for outliers, which would heavily distort the following metrics. To assess the reproducibility of the perfusion values within the grey matter and white matter masks, the within-subject coefficient of variation (wsCV) was calculated using the logarithmic method \cite{Bland1996log,Bland2006} and the intraclass correlation coefficient (ICC) for test-retest data \cite{McGraw1996,Koo2016} was calculated using available Matlab Code \cite{Salarian2026}. As wsCV and ICC require singular values, not images, they were calculated on the mean perfusion for grey matter and white matter. To better capture signal variations in the grey and white matter, a voxel-based comparison using root mean squared error (RMSE) was also performed between two measurements of the same subject. For this, the CBF maps were registered to standard space using FSL.  
\section{Results}
\begin{figure}
		\centering
		\includegraphics[width=400pt]{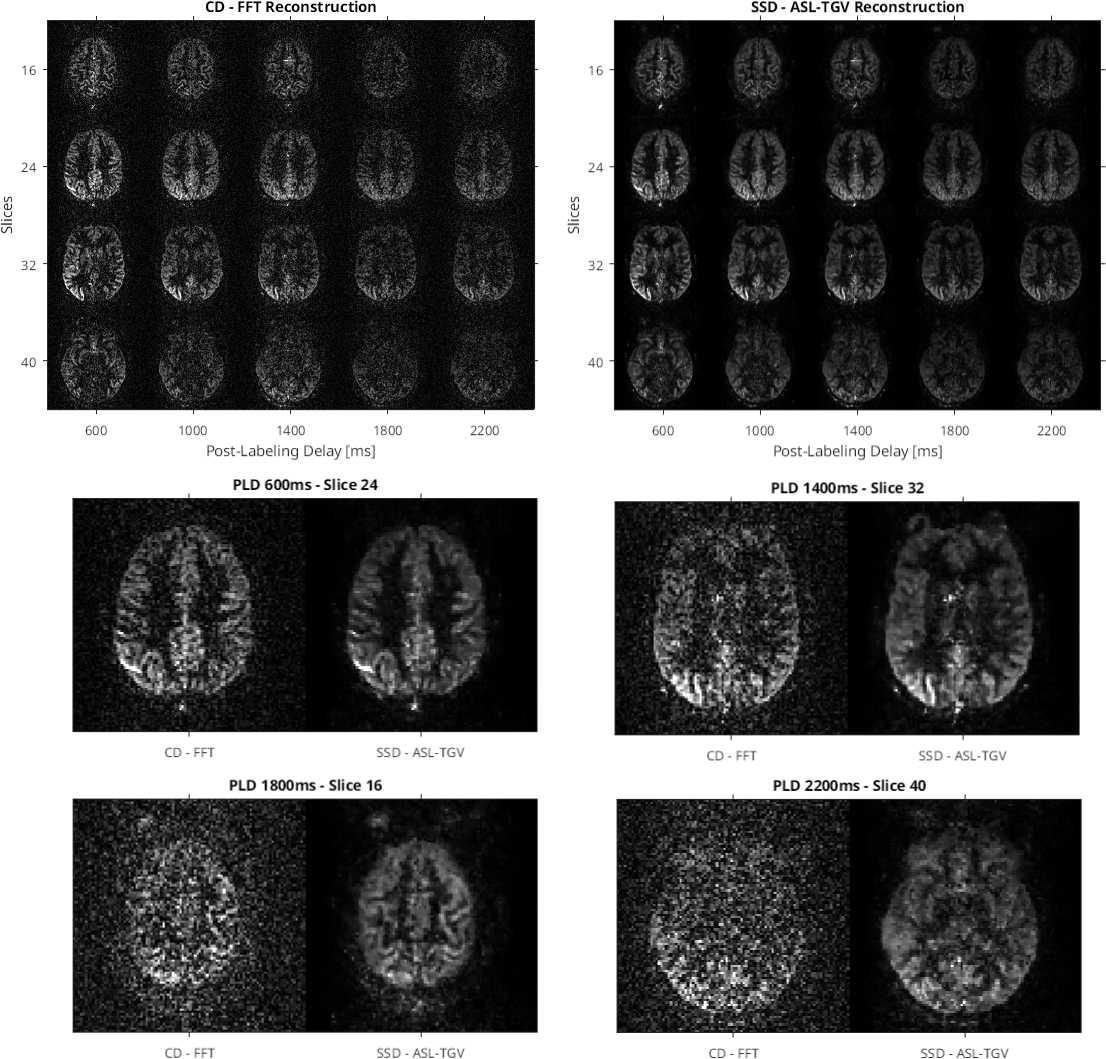}
		\caption{These are the images created with the two different reconstructions for the first visit of subject 1. The top row shows four selected slices for each PLD for both reconstructions. Below are direct comparisons of the two reconstructions for the selected slices at one of the PLDs. \label{fig_5PLD}}
\end{figure}

The full set of reconstructed images for the first visit of subject 1 for the combined and single-shot dataset are presented in figure \ref{fig_5PLD}. The eight acquisitions at each PLD in the single-shot dataset are averaged for the figure. 
In addition to the full sets, a larger comparison of a few indiviudual slices is included to better compare directly. 
Looking at the full sets, while it is recognizable that both sets represent the same data from the same subject, some clear differences remain. 
The ASL-TGV  has significantly reduced noise and more clearly resolved structures especially for the later PLDs and the outermost slices. 

The individual slices at selected PLDs from subject 1 are representative for the dataset and display the differences between the reconstructions more clearly. 
For all slices and PLDs the ASL-TGV reconstruction is less noisy and more clearly resolves the structures in the PWIs. 
The improvement in quality increases for the later PLDs and the outermost slices. 
For the center slice, slice 24, at PLD 600ms, the FFT reconstruction shows more noise, but the two images are recognizable as depicting the same subject. 
Examples like slice 40 at PLD 2200ms, the image of the FFT reconstruction is dominated by noise, while the ASL-TGV produces a good quality depiction of the slice. 

The adaptively denoised images can be seen in the appendix in \ref{fig_5PLD_denoised}. While the PWIs from the ASL-TGV reconstruction barely changed, the noise levels in the FFT reconstruction are clearly reduced. For the central slices with better SNR after reconstruction, the denoised images from the two different reconstructions look almost identical. In the slices from the later PDLs in the outermost parts of the brain, that were much more impacted by noise, there is a clear difference between the two reconstructions. 

\begin{center}
	\begin{figure}
		\includegraphics[width=400pt]{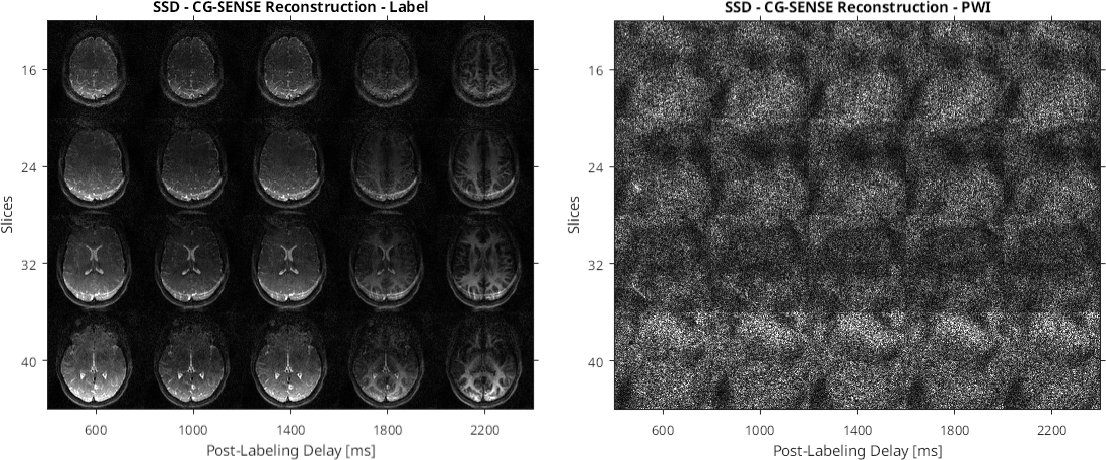}
		\caption{The averaged label images and PWIs reconstructed from the single-shot dataset by a conventional CG-SENSE reconstruction. While the label images can be reconstructed reasonably well, the perfusion information is completely lost due to the SNR penalty of the undersampling.  \label{fig_CG-SENSE}}
	\end{figure}
\end{center}

The single-shot dataset was also reconstructed using a conventional CG-SENSE approach. Figure \ref{fig_CG-SENSE} reveals that the label images are reconstructed reasonably well showing the different contrasts due to the different timings of the background suppression. However, the PWIs are entirely noise, no perfusion information can be recovered from those images. For these reasons, these reconstructions will no longer be considered for the further analysis. 

\begin{center}
	\begin{figure}
		\includegraphics[width=400pt]{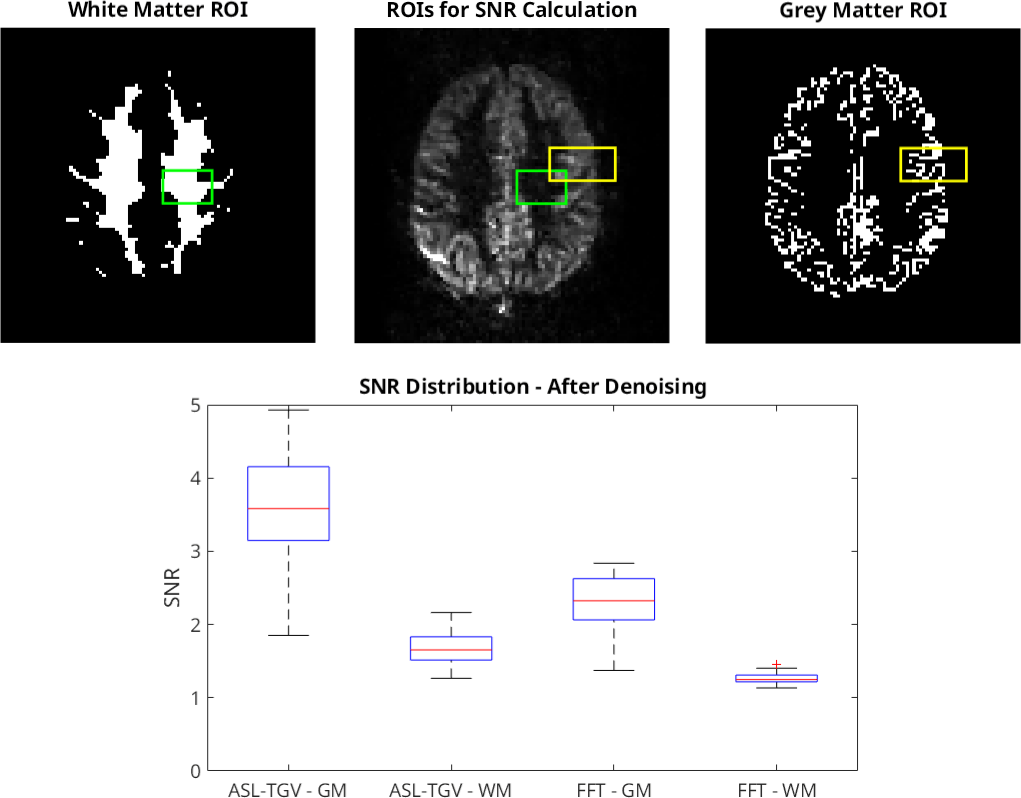}
		\caption{The SNR and the two regions of interest of the PWIs for the ASL-TGV reconstruction and FFT reconstruction after denoising. For ASL-TGV, the SNR is clearly better in both grey matter and white matter. It is 62\% and 35\% better when averaged over all subjects and all PLDs. \label{fig_SNR}}
	\end{figure}
\end{center}

The visual difference of the PWIs between the FFT and ASL-TGV reconstructions is confirmed by the SNR estimation, which is detailed in figure \ref{fig_SNR}. 
For the estimation of the grey matter metrics, an average of 180 pixels were considered, while for the white matter 424 pixels were used on average. 
The SNR is significantly higher for ASL-TGV for both grey and white matter. 
A notable result is that the SNR in white matter in PWIs from the FFT reconstruction appears to be practically constant for all subjects. 
Averaging over all subjects and PLDs, ASL-TGV has a 62\% higher SNR in the grey matter ROI and a 35\% higher value in the white matter ROI. Looking at the values per PLD, the improvement becomes more pronounced for the longer PLDs. For the first PLD, the SNR increase is 40\% and 30\% for grey and white matter. For the last PLD, it is 80\% and 46\%.  

\begin{figure}
	\centering
	\includegraphics[width=250pt]{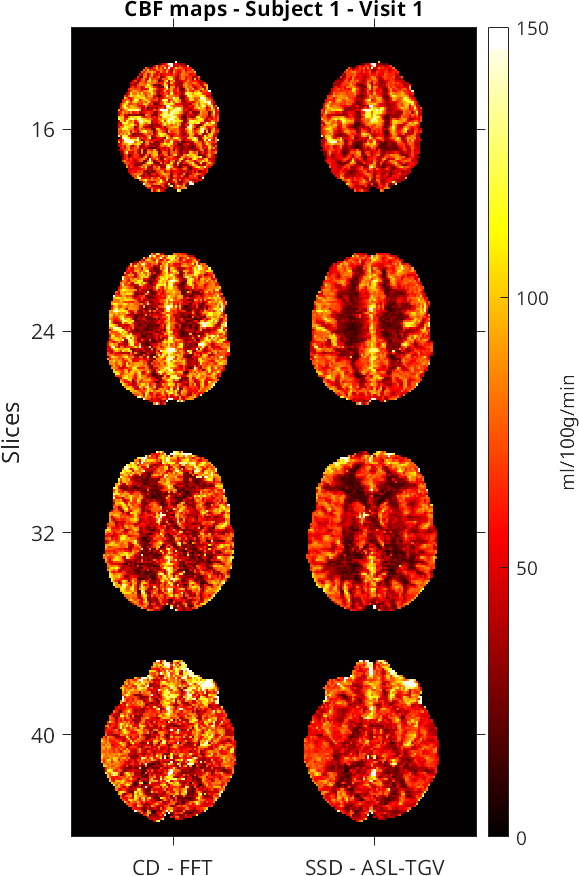}
	\caption{The CBF maps calculated based on the FFT reconstruction of the combined dataset and ASL-TGV reconstruction based on the single-shot dataset. The averages of grey and white matter perfusion over all subjects show that ASL-TGV produces slightly lower perfusion estimates overall, 7\% decrease for white matter, 5\% for grey matter. However, the respective ratios of grey and white matter are practically equal. }\label{fig_CBF_maps1}
\end{figure}

The CBF maps from the two sets of images can be seen in figure \ref{fig_CBF_maps1}. The differences in the PWIs translate to small differences in estimated CBF values. Over all subjects, the perfusion is calculated to be 7\% lower in white matter and 5\% lower in grey matter for the estimates derived from ASL-TGV PWIs. The ratio of white to grey matter perfusion is approximately equal between the two methods. 

\begin{table}[ht]
	\centering
	\begin{tabular}{c||cc} 
		\multicolumn{3}{c}{Average Perfusion [ml/100g/min]}\\
		Based on & \multicolumn{2}{c|}{All Subjects}\\
		Recon & FFT Recon &  ASL-TGV Recon\\
		\hline
		Grey Matter & 69 & 66\\
		White Matter & 47 & 44\\
		\hline
	\end{tabular}
	\caption{The mean perfusion values for white matter and grey matter regions averaged over all patients. }\label{tab_perfusion}
\end{table}

\begin{figure}
	\centering
	\includegraphics[width=300pt]{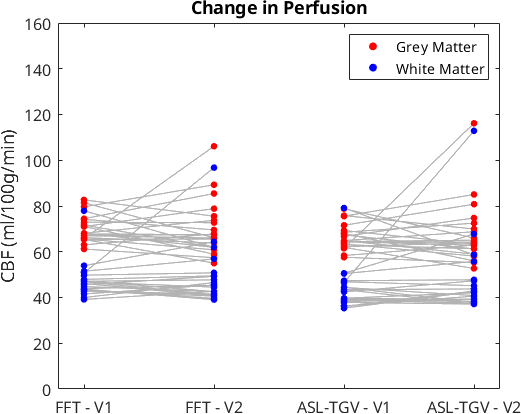}
	\caption{The change in grey and white matter perfusion for each subject and each reconstruction method. V1 and V2 stand for Visit 1 and Visit 2 of each subject. }\label{fig_slope_chart}
\end{figure}

The actual change in perfusion for each patient in grey and white matter for the two reconstruction methods is depicted in figure \ref{fig_slope_chart}. 

Before quantifying the test-retest reliability, subject 21 was excluded from the calculations. Both the images and the CBF maps can be found in the appendix in figure \ref{fig_sub21}. The images show strong artifacts which translate into highly irregular CBF estimates, which are also apparent in figure \ref{fig_slope_chart}. These values had a large impact on the following metrics and are thus excluded from the RMSE, ICC and wsCV calculations. 

\begin{figure}
	\centering
	\includegraphics[width=250pt]{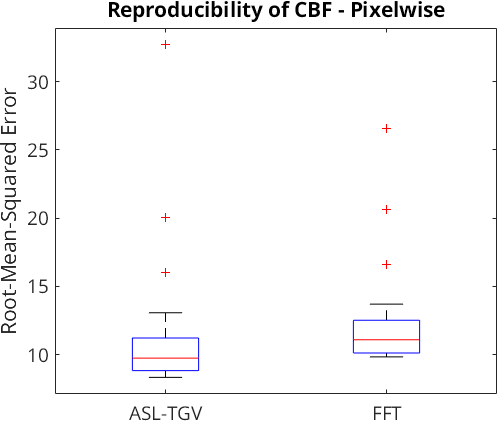}
	\caption{This figure gives an overview of the pixelwise test-retest reliability of the FFT reconstruction on the combined dataset and the ASL-TGV reconstruction on the single-shot dataset.  The RMSE averaged over all subjects except the outlier subject 21 suggests that the test-retest reliability is better for the maps derived from the ASL-TGV images. The average RMSE values are 10.6 for the ASL-TGV reconstruction and 11.8 for the FFT reconstruction.  }\label{fig_region_barplot_02}
\end{figure}

Figure \ref{fig_region_barplot_02} provides the test-retest analysis based on the RMSE between the CBF maps of the first and second visit of each subject. The average over all considered subjects shows that ASL-TGV has an approximately 10\% better RMSE. 

Table \ref{tab_ICC} shows the ICC and wsCV values for grey and white matter perfusion. 

\begin{table}[ht]
	\centering
	\begin{tabular}{c||cc|cc}
		\multicolumn{5}{c}{Test-Retest Metrics}\\
		Metric & \multicolumn{2}{c|}{ICC} & \multicolumn{2}{c}{wsCV}\\
		Recon & FFT & ASL-TGV & FFT & ASL-TGV  \\
		\hline
		Grey Matter & 0.54 & 0.56 & 7.6\% & 7.2\% \\
		White Matter & 0.73 & 0.68 & 7.6\% & 9.7\% \\
		\hline
	\end{tabular}
	\caption{The ICC and wsCV calculated for grey and white matter perfusion derived from both denoised reconstructions. The optimal ICC is 1, the optimal wsCV is 0\%. Subject 21 was excluded as an outlier (see Appendix). All ICC values reveal only moderate reproducibility  ASL-TGV is better than the FFT reconstruction in grey matter and the FFT reconstruction is better in white matter. } \label{tab_ICC}
\end{table}

\newpage
 
\section{Discussion}
The overall goal of this work is to extend the ASL-TGV reconstruction to multi-delay ASL and show its advantages against using only basic reconstruction techniques like an FFT or SENSE/GRAPPA in both image quality and reproducibility. As there are no general recommendations for specialized ASL image reconstruction to compare against, the reference reconstruction was chosen to be the FFT reconstruction. This dataset is particularly well-suited to make this comparison as the fully sampled combined dataset and the highly undersampled single shot dataset are based on the exact same raw data. This enables a direct comparison of the two methods. It is impossible to fully determine which set of images most truly represents the perfusion of the subject, but there are several factors showing the advantages of the ASL-TGV reconstruction. 

Figure \ref{fig_5PLD}, the set of reconstructed PWIs from the test-retest dataset, shows that advantage directly. The general decrease in noise from ASL-TGV is expected from a regularized reconstruction, but the effect is so pronounced that images that are dominated by noise in the FFT reconstruction reveal the underlying structures in the ASL-TGV reconstruction. This new reconstruction not only produces an improvement in image quality for data with inherently reasonable SNR like the example slice 24 at PLD 600ms, it even recovers the image in cases where the FFT reconstruction only produces almost unreadable images like slice 40 at PLD 2200ms. 

As it is recommended to use at least one denoising technique on PWIs \cite{Hernandez-Garcia2022}, an adaptive 3D Wiener Filter was applied to all images. The filter clearly improves the PWIs from the FFT reconstruction. For the images with little noise, it even improves them to a comparable quality of the ASL-TGV reconstruction. However, for the images that were originally dominated by noise, the denoised PWIs are blotchy despite using an adaptive filter. The changes to the ASL-TGV PWIs by the filter are almost imperceptible. 

The images of the single shot dataset using a conventional CG-SENSE reconstruction in figure \ref{fig_CG-SENSE} show that the highly undersampled ASL data cannot be reconstructed without advanced image reconstruction. While image denoising techniques have made great progress in general and for ASL in particular, these approaches would not be able to recover the perfusion information from those images. The data must be reconstructed with ASL-TGV or something equally powerful to use the data in a single-shot manner. 

The SNR improvement of the ASL-TGV images over the denoised FFT reconstruction is large with 62\% in the grey matter ROI and 35\% in the white matter ROI. It must be noted that the location of the ROIs in a central slice were chosen in a way that gave the FFT reconstruction the best possible result. Using the outermost slices the FFT reconstruction would have even lower SNR or made SNR calculation fully impossible. 

Estimating CBF maps shows that the quantification reflect the results from the PWIs, as the maps derived from the ASL-TGV PWIs are noticably less noisy. The estimated perfusion values differ slightly between the two reconstruction methods, but that small bias in CBF is expected from the TGV regularization \cite{Spann2020}. Still, as the ratio between grey and white matter perfusion is approximately constant and the absolute values only differ by 3 ml/100g/min respectively, the overall information remains similar. The mean grey matter perfusion is within the expected range compared to literature, but the white matter perfusion estimates from both reconstructions are higher than typical literature values \cite{Farahani2025,Juttukonda2021}. 

The results from the test-retest reliability demonstrate that the ASL-TGV reconstruction produces more consistent images for the given dataset. 
The pixelwise comparison via the RMSE reveals a 10\% decrease in difference for the ASL-TGV reconstruction compared to the FFT reconstruction. 
Using ICC and wsCV metrics on perfusion reveals a more nuanced aspect. 
ICC is better in grey matter for ASL-TGV, but worse in white matter. For wsCV, the relations are identical. While this indicates better test-retest repeatability in white matter for the FFT reconstruction, other results from this work cast a doubt on that. The very low and practically constant SNR over all subjects in white matter for the FFT reconstructions suggests that the more repeatable perfusion might actually come from fitting and averaging the noise instead of fitting a repeatable signal. 

Comparing ICC and wsCV to other test-retest studies, the wsCV is within the typical range, while the ICC is on the lower end of typical ASL perfusion test-retest reliability studies. \cite{Petersen2010,Gevers2011,Chen2011_testretest,Jain2012,Mutsaerts2014,Mutsaerts2015,Ssali2016,McGuire2017,Melzer2020,Cohen2020}. A possible cause may be the combination of a high-resolution paired with a highly accelerated acquisition scheme. Another possibility is the use of a pediatric dataset as another test-retest study on pediatric data produced similar ICC values \cite{Jain2012}. 

The ICC and wsCV metrics calculated are improved compared to the ones originally calculated in Shou et al \cite{Shou2025} for this dataset for ASL-TGV. There, the previous SD-ASL version of ASL-TGV developed by Spann et al was used with different reconstruction parameters and a different custom post-processing protocol. While ASL is a quantitative method, it is important to remember the large impact the choices of post-processing parameters and reconstruction have on the ultimate results. This is made obvious in the comparison study by Paschoal et al \cite{Paschoal2024}. 

Concluding the discussion about test-retest metrics, these discrete values of improvement of the ASL-TGV approach against the FFT reconstruction are not to be taken as a definitive measure. The perfusion cannot be assumed to be identical for each subject for both visits, so the RMSE or wsCV would actually never be zero or the ICC would never be 1 even with a perfect acquisition and reconstruction. While this test-retest scheme cannot be used to declare a specific percentage improvement of one reconstruction to the other, there is an improvement to most ASL-TGV derived values. 

In addition to these advantages of ASL-TGV in the PWIs, CBF maps and test-retest reliability, it also enables reconstructing images from single-shot acquisition, which enhances the temporal resolution and inherently leads to more motion robustness.

\section{Conclusions}\label{sec_conclusion}
The proposed MD-ASL reconstruction enables high resolution, single-shot ASL acquisitions and improves both SNR and reproducibility. It additionally complements all existing advances in both data acquisition and post-processing. As ASL-TGV uses no particular ASL model, does not require any training and only uses the concept of the difference image in its formulation, it can be applied to all labeling types and any type of k-space trajectory. It is available open-source in BART and the reconstructed images can be plugged into any existing post-processing pipeline. The available higher temporal resolution may enable more fine-tuned motion correction in the future.  

\section*{Conflict of Interest}
The authors declare no competing interests.

\section*{Data Availability Statement}
An exemplary dataset of one subject is available together with a script to reconstruct the images using BART via \doi{10.5281/zenodo.22807527}.\\
The BART code together with installation instructions can be found here: \\ \url{https://mrirecon.codeberg.page/installation.html}.\\
A detailed explanation of the BART command to call this reconstruction can be found in the appendix. \\
Alternatively, the same example reconstruction can be run on a Google Colab: \\
\url{https://colab.research.google.com/drive/105OPnQ8VsgHxmF11KWhcULUNOMre5pnh?usp=sharing}

\section*{Acknowledgements}

\printfunding

\printbibliography

\appendix

\section{Using ASL-TGV in BART\label{app1}}
\subsection{How to Call ASL-TGV with BART\label{appBARTrecon}}
To use the presented reconstruction approach in BART, the kspace file must be prepared in the appropriate format. 
In addition to using the BART .cfl and .hdr file format, this means saving the data matrix into the correct dimensions. For Cartesian data, the $k_x$, $k_y$ and $k_z$ data must go into dimensions 0, 1 and 2. 
The coils are in dimension 3. 
The control/label pairs go in dimension 8, which therefore must have length 2. 
Finally, the different delays and averages go into dimensions 10 and 11, respectively. 
If the data is non-Cartesian, the file must follow the BART non-Cartesian data format in the first three dimension and one needs an additional trajectory file. 
Thus, given a k-space file \textit{kspace} and a coil sensitivity file \textit{sens}, the command may look as follows. 
\begin{verbatim}
	bart pics --asl -g -i 500 -S -R G:$(bart bitmask 0 1 2 10 11):0:1.0 
	--tvscales 1:1:1:1:5 --theta 1:5 kspace sens result
\end{verbatim}
The options in this command do the following: 
\begin{itemize}
	\item \textit{- -asl} applies the TV or TGV regularization on the PWI in addition to the control-label images.
	\item \textit{-g} (optional) enables use of GPU, if supported on this system. This will significantly speed up the reconstruction. 
	\item \textit{-i} (optional) sets the number of iterations for the algorithm. For heavily undersampled data, the default number of iterations may not suffice and must be set higher as shown here. 
	\item -S In BART the data is automatically scaled during reconstruction. The -S flag ensures that the data is scaled back after the reconstruction. This is important for quantification to ensure that all reconstructions are scaled equally. 
	\item -R selects the reconstruction type, dimensions and strength. \textit{T} selects Total Variation, \textit{G} selects Total Generalized Variation 2nd order, the next number selects the dimensions on which the regularization is applied via a bart bitmask (in this case the three spatial dimensions, the label and average dimension), the next number is the joint threshold flags and can usually be kept at $0$ and the final number is the strength of the regularization ($\lambda$ in the reconstruction equation)
	\item \textit{- -tvscales} selects how the gradients of T(G)V are weighted ($\beta$ in the reconstruction equation). The ones in the spatial dimension can be adapted to the resolution (isotropic resolution suggests equal weighting for spatial weights) and the temporal dimensions can be weighted according to meaning (averages can be weighted highly, dynamic dimensions can be weighted lower). 
	\item \textit{- -theta} selects the weighting between the regularization of the control/label images and the perfusion weighted images ($\theta_1$ and $\theta_2$ in the reconstruction equation). The second number weights the PWI and we recommend to make it larger due to the smaller magnitude of the PWI and since it is the final target of the reconstruction. 
\end{itemize}

\subsection{Example Data - Subject 1, Visit 1}
As written in the data availability statement, the k-space data of subject 1, visit 1 is available via \doi{10.5281/zenodo.22807527} together with a script that performs the reconstruction. This requires BART installed, preferably with GPU support. Alternatively, a Google Colab script is available, which installs BART, downloads the data, reconstructs the image including coil sensitivity estimation and plots an image.\\ \url{https://colab.research.google.com/drive/105OPnQ8VsgHxmF11KWhcULUNOMre5pnh?usp=sharing} \\This example data and script can be used as a blueprint for your own ASL reconstructions. 

\newpage

\section{Additional Data}
\subsection{Subject 1 - PWIs - Denoised}
\begin{figure}[h!]
	\centering
	\includegraphics[width=400pt]{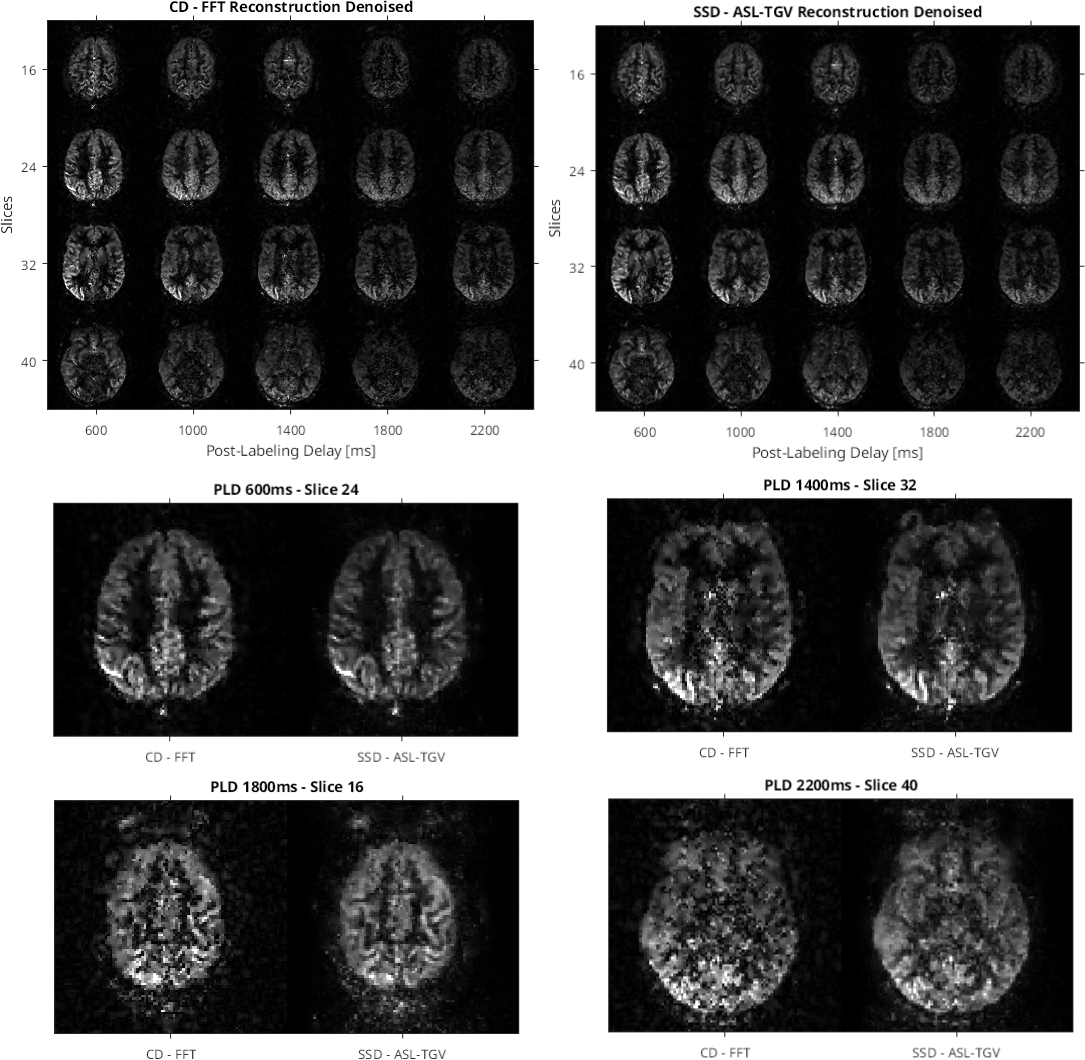}
	\caption{These are the denoised images created with the two different reconstructions for the first visit of subject 1. The top row shows four selected slices for each PLD for both reconstructions. Below are direct comparisons of the two reconstructions for the selected slices at one of the PLDs.  \label{fig_5PLD_denoised}}
\end{figure}

\subsection{Subject 21 - PWIs and CBF - Outlier}
\begin{figure}[h!]
	\centering
	\includegraphics[width=400pt]{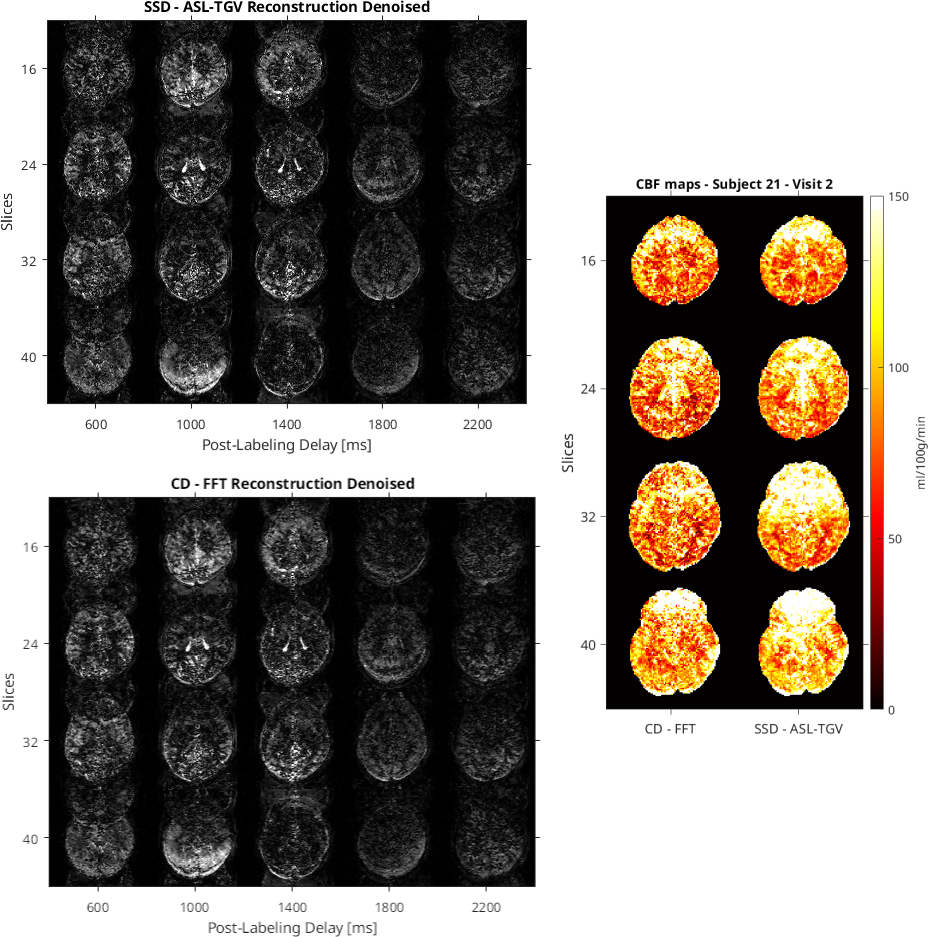}
	\caption{The images and CBF map of visit 2 of subject 21. The artifacts in the image produce a highly irregular CBF map. This subject was thus excluded from the test-retest evaluation.  \label{fig_sub21}}
\end{figure} 
\clearpage


\end{document}